\documentclass[aps,twocolumn,amsmath,amssymb,showpacs,showkeys,floatfix]{revtex4}
\usepackage{graphicx}
\usepackage{amssymb} 

\begin{document}




\title{Pygmy dipole resonance: collective features and symmetry energy effects}

\author{V.Baran $^{1}$, B. Frecus$^{1}$, M. Colonna$^{2}$, M. Di Toro$^{2,3}$}
\affiliation{$^{1}$ Physics Faculty, University of Bucharest, Romania},
\affiliation{$^{2}$ Laboratori Nazionali del Sud INFN, I-95123 Catania, Italy}
\affiliation{$^{3}$ Physics and Astronomy Dept., University of Catania, Italy}

\begin{abstract}
 A very important open question related to the pygmy dipole resonance is about
its quite elusive collective nature. In this paper, within a
harmonic oscillator shell model, generalizing an approach introduced
by Brink, we first identify the dipole normal modes in neutron rich
nuclei and derive 
the energy weighted sum rule
exhausted by the pygmy dipole resonance. Then solving numerically
the self-consistent Landau-Vlasov kinetic equations for neutrons and
protons with specific initial conditions, we explore the structure
of the different dipole vibrations in the $^{132}Sn$ system and
investigate their dependence on the symmetry energy. We evidence the
existence of a distinctive collective isoscalar-like mode with an
energy well below the Giant Dipole Resonance (GDR), very weakly dependent
on the isovector part of the nuclear effective interaction. At
variance the corresponding strength is rather sensitive to the
behavior of the symmetry energy below saturation, which rules the
number of excess neutrons in the nuclear surface.
\end{abstract}



\maketitle


 One of the important tasks in many-body physics is to understand the emergence of the
collective features as well as their structure in terms of the individual motion of the
constituents. The steady progress of experimental methods of investigation opens now
the possibility to study very neutron rich nuclei, beyond the  limits of stability.
The goal is to have a unified picture of the evolution of various nuclear properties with mass and
isospin and to test the validity of our theoretical understanding over an extended domain
of analysis.

  New exotic collective excitations show up when one moves away from the valley of
stability \cite{paar2007}. Their experimental characterization and theoretical
description is a challenge for modern nuclear physics. Recent experiments provided
several evidences  about their existence but the available information is still
incomplete and their nature is a matter of debate.

   An interesting exotic mode is the Pygmy Dipole Resonance (PDR) which was observed as
an unusually large concentration of the dipole response at energies
clearly below the values associated with the 
GDR. The latter is one of the most prominent and robust collective
motions, present in all nuclei, whose centroid position varies, for
medium-heavy nuclei, as $80 A^{-1/3} MeV$.
Adrich et al. \cite{adrich2005} reported the observation of a resonant-like shape distribution
with a pronounced peak around $10MeV$ in $^{130}Sn$ and $^{132}Sn$ isotopes.
A concentration of dipole excitations near and below the particle emission threshold was
also observed in stable Sn nuclei, a systematics of
PDR in these systems being presented in \cite{ozel2007}. It was
concluded that the strongest transitions locate at energies between
$5$ and $8.5 MeV$ and a sizable fraction of the Energy-Weighted Sum
Rule (EWSR) is exhausted by these states. From a comparison of the
available data for stable and unstable $Sn$ isotopes a correlation
between the fraction of pygmy strength and isospin asymmetry was
noted \cite{klimkiewicz2007}. In general the exhausted sum-rule
increases with the proton-to-neutron asymmetry. This behavior was
related to the symmetry energy properties below saturation and
therefore connected to the size of the neutron skin
\cite{yoshida2004,piekarewicz2006,carbone2010}. However other
theoretical analyses suggest a weak connection between the PDR and
skin thickness \cite{reinhard2010}.

 In spite of the theoretical progress in the interpretation of
this mode within phenomenological studies based on hydrodynamical equations
 \cite{mohan1971,bastrukov2008}, non-relativistic microscopic
approaches using Random Phase Approximation (RPA) with various
effective interactions \cite{tsoneva2008,co2009,yoshida2009} or
relativistic quasi-particle RPA
\cite{vretenar2001,litvinova2008}, and new experimental
information \cite{savran2008,
wieland2010,tonchev2010,makinaga2010}, a number of critical
questions concerning the nature of the PDR still remains. This
includes the macroscopic picture of neutron and proton vibrations,
the exact location of the PDR excitation energy and the degree of
collectivity of the low-energy dipole states,
the role of the symmetry energy \cite{paar2010}.
Some microscopic studies predict a large fragmentation of the GDR strength and
the absence of collective states in the low-lying region in $^{132}Sn$ \cite{sarchi2004}.

The purpose of this letter is to address the important issue related
to the collective nature of PDR. In the first part, within the
Harmonic Oscillator Shell Model (HOSM) for neutron rich nuclei,
we show that the coordinates associated with the neutron excess
vibration against the core, and to the dipole core mode respectively, are separable
and derive 
the EWSR exhausted by each of them.
Then we adopt a description based on the Fermi liquid theory
with effective interactions and investigate the dynamics and the interplay
between the dipole modes identified in HOSM. This self-consistent
treatment allows us to inquire on the role of the symmetry energy
and its density dependence upon the dipole response.

 In a seminal paper \cite{brink1957}, Brink has shown that for a
system of $A=N+Z$ nucleons moving in a harmonic oscillator well with the  Hamiltonian
$ \displaystyle H_{sm} = \sum_{i=1}^{A} \frac{\vec{p}_i^2}{2 m}+
\frac{K}{2} \sum_{i=1}^{A} \vec{r}_i^2~, $
it is possible to perform a separation in four independent parts
$\displaystyle H_{sm} = H_{n~int}+H_{p~int}+H_{CM}+H_{D} $.
The first two terms
determine the internal motion of protons and neutrons
respectively, depending only on proton-proton and neutron-neutron relative coordinates.
The Hamiltonian
$ \displaystyle H_{CM} = \frac{1}{2Am} \vec{P}_{CM}^2+\frac{KA}{2}\vec{R}_{CM}^2 $
characterizes the nucleus center of mass (CM) motion, while
$ \displaystyle H_{D} = \frac{A}{2mNZ} \vec{P}^2+\frac{K NZ}{2A}\vec{X}^2,$
describes a Goldhaber-Teller (G-T) \cite{goldhaber1948} vibration of protons against neutrons.
The oscillator constant $K$ can be determined by fitting the nuclear size \cite{bohr1998}.
In the expressions above
$\vec{X}$  and $\vec{R}_{CM}$ denote the neutron-proton relative coordinate
and the center of mass position, and we have
introduced the conjugate momenta 
$\displaystyle \vec{P}=\frac{N Z}{A}(\frac{1}{Z} \vec{P}_Z- \frac{1}{N} \vec{P}_N) $
and
$\displaystyle \vec{P}_{CM}=\vec{P}_{Z}+\vec{P}_{N}$,
where $\vec{P}_{Z}$ ($\vec{P}_{N}$) are proton (neutron) total momenta.
Correspondingly, the eigenstates of the nucleus are
represented as a product of four wave functions $\displaystyle \Psi=\psi_{n~int}
\chi_{p~int} \alpha (\vec{R}_{CM}) \beta(\vec{X})$,
simultaneous eigenvectors of the four Hamiltonians constructed above.
For an $E1$ absorption a G-T collective motion  with a specific linear
combination of single particle excitations is produced
and the wave function $\beta(\vec{X})$ is changing from the ground state
to the one GDR phonon state.
Denoting by $E_i$ the energy eigenvalues of the system 
and by $D$ the dipole operator ,
with the help of the Thomas-Runke-Kuhn (TRK) sum rule, the total absorption cross section is given by:
$\displaystyle \sigma_D = \int_0^\infty \sigma(E) dE=
\frac{4 \pi^2 e^2}{\hbar c} \sum_i E_i  |\langle i |D| 0 \rangle|^2
=\frac{4 \pi^2 e^2}{\hbar c} \frac{1}{2} \langle 0 |[D,[H_{sm},D]] |0 \rangle
= 60 \frac{NZ}{A} mb \cdot MeV $.
 
Now let us turn to the physical situation, corresponding to the case of very neutron-rich nuclei,
where the system is conveniently described in terms of a bound core containing
all protons and $N_c$ neutrons, plus some (less bound) excess neutrons $N_e$.
Thus the total neutron number $N$ is split into the sum $N = N_c + N_e$ and
we denote by $A_c = Z + N_c$ the number of nucleons contained in the core.
In this case we have worked out an exact separation of the HOSM
 Hamiltonian in a sum of six independent (commuting) quantities:
$\displaystyle H_{sm}= H_{n_c~int}+H_{p_c~int}+H_{e~int}+ H_{CM}+H_{c} + H_{y}$.
The first three terms
contain only relative coordinates and momenta among nucleons of each ensemble,
i.e. core neutrons, core protons
and excess neutrons and, as before, describe their internal motion.
$\displaystyle H_{c}=\frac{A_c}{2Z~N_c m}\vec{P}_{c}^2+\frac{K N_c Z}{2A_c}\vec{X}_{c}^2,$
characterizes the core dipole vibration, while the relative
motion of the excess neutrons against the core, usually associated with the pygmy mode, is determined
by  $ \displaystyle H_{y}=\frac{A}{2A_c N_e m}\vec{P}_{y}^2 + \frac{K N_e A_c}{2 A}\vec{Y}^2$.
Here $\displaystyle \vec{X}_{c} $ denotes the distance between neutron and proton centers of
mass in the core, while
$\displaystyle \vec{Y}$
is the distance between the core center of mass and the center of mass of the excess neutrons.
The corresponding canonically conjugate momenta are
$\displaystyle \vec{P}_{c}=\frac{N_c Z}{A_c}(\frac{1}{Z} \vec{P}_{Z}-\frac{1}{N_c}\vec{P}_{N_c})$,
and $\displaystyle \vec{P}_{y}=\frac{N_e A_c}{A}(\frac{1}{A_c}(\vec{P}_{Z}+\vec{P}_{N_c})-\frac{1}{N_e}
\vec{P}_{N_e})$.
The eigenstates of  $H_{c}$ and
$H_{y}$ are describing two independent collective excitations and
both of them will contribute to the dipole response since
the total dipole momentum can be expressed as:
$\displaystyle \vec{D}=\frac{N Z}{A} \vec{X}= 
\frac{Z~ N_c}{A_c} \vec{X}_{c} + \frac{Z~ N_e}{A} \vec{Y} \equiv \vec{D}_c+\vec{D}_y.$
 In this picture the PDR results in a collective motion of
G-T type with the excess neutrons oscillating against the core.
The $E1$ absorption leads also to the change of the wave function
associated with the coordinate $\vec Y$. The total cross section for the
PDR is:
$\displaystyle \sigma_y = \frac{4 \pi^2 e^2}{\hbar c} \frac{1}{2} \langle 0 |[D_y,[H_{sm},D_y]]|0 \rangle =
\frac{N_e Z}{N A_c} \sigma_D.$
This shows that a fraction $ \displaystyle f_y=\frac{N_e Z}{N A_c}$
of the EWSR is exhausted by the pygmy mode. It is worth to mention
that this result is consistent with the molecular sum rule introduced by
Alhassid et al. \cite{alhassid1982}. For the tin isotope $^{132}Sn$,
if the excess neutrons were simply defined as the difference between
neutron and proton numbers, i.e. $N_e=32$,
one would expect $f_y = 19.5 \%$. This is
greater than the value estimated experimentally, which is around $5
\%$. A possible explanation for this difference is that 
only a part of the excess neutrons, $N_{y}$, with $N_{y}<N_e$,
contribute to PDR, the rest being still bound to the core
\cite{ncore}.

 Therefore it is important to test this assumption within a more
sophisticated analysis of the dipole response. Indeed, a more
accurate picture of the GDR in nuclei corresponds to an admixture of
G-T and Stenweidel-Jensen (S-J) vibrations. The latter, in
symmetric nuclear matter, is a volume type oscillation of the
isovector density $\rho_i= \rho_n - \rho_p$  keeping the total
density  $\rho=\rho_n+\rho_p$ constant \cite{steinwedel1950}.
A microscopic, self-consistent study of the collective features
and of the role of the nuclear effective interaction
upon the PDR can be performed within the Landau theory of Fermi liquids.
This is based on two coupled Landau-Vlasov kinetic equations for neutron and proton one-body
distribution functions $f_q(\vec{r},\vec{p},t)$ with $q=n,p$:
\begin{equation}
\frac{\partial f_q}{\partial t}+\frac{\bf p}{m}\frac{\partial f_q}{\partial {\bf r}}-
\frac{\partial U_q}{\partial {\bf r}}\frac{\partial f_q}{\partial {\bf p}}=I_{coll}[f] ,
\label{vlasov}
\end{equation}
and was applied  quite successfully in describing various features of the GDR,
including pre-equilibrium dipole excitation in fusion reactions \cite{baran1996}.
Within a linear response approach, it was also considered to investigate
properties of the PDR \cite{abrosimov2009}.
However, it should be noticed that 
within such a semi-classical description
shell effects are absent, certainly important in shaping the fine structure
of the dipole response \cite{maza2012}.
We neglect here the two-body collisions effects and hence the main ingredient of
the dynamics is the nuclear mean-field, for which we consider a Skyrme-like ($SKM^*$) parametrization
$\displaystyle U_{q} = A\frac{\rho}{\rho_0}+B(\frac{\rho}{\rho_0})^{\alpha+1} + C(\rho)
\frac{\rho_n-\rho_p}{\rho_0}\tau_q
+\frac{1}{2} \frac{\partial C}{\partial \rho} \frac{(\rho_n-\rho_p)^2}{\rho_0}$,
where $\tau_q = +1 (-1)$ for $q=n (p)$ and $\rho_0$ denotes the saturation density.
The saturation properties of symmetric nuclear matter are reproduced
with the values of the coefficients
$A=-356 MeV$, $B=303 MeV$, $\alpha=1/6$,
leading to a compressibility modulus $K=200 MeV$. For the isovector sector we employed
three different parameterizations of $C(\rho)$ with the density: the asysoft,
the asystiff and asysuperstiff respectively, see \cite{baran2005} for a detailed description.
The value of the symmetry energy,
$\displaystyle E_{sym}/A = {\epsilon_F \over 3}+{C(\rho) \over 2}{\rho \over \rho_0}$,
at saturation, as well as
the slope parameter, $\displaystyle L = 3 \rho_0 \frac{d E_{sym}/A}{d \rho} |_{\rho=\rho_0}$,
are reported in Table \ref{table1} for each of these asy-EoS. Just below the saturation density
the asysoft mean field has a weak variation with density while the asysuperstiff shows
a rapid decrease. 
Then, due to surface
contributions to the collective oscillations, we expect to see
some differences in the energy position of the dipole response of the system.

\begin{figure}
\begin{center}
\includegraphics*[scale=0.27]{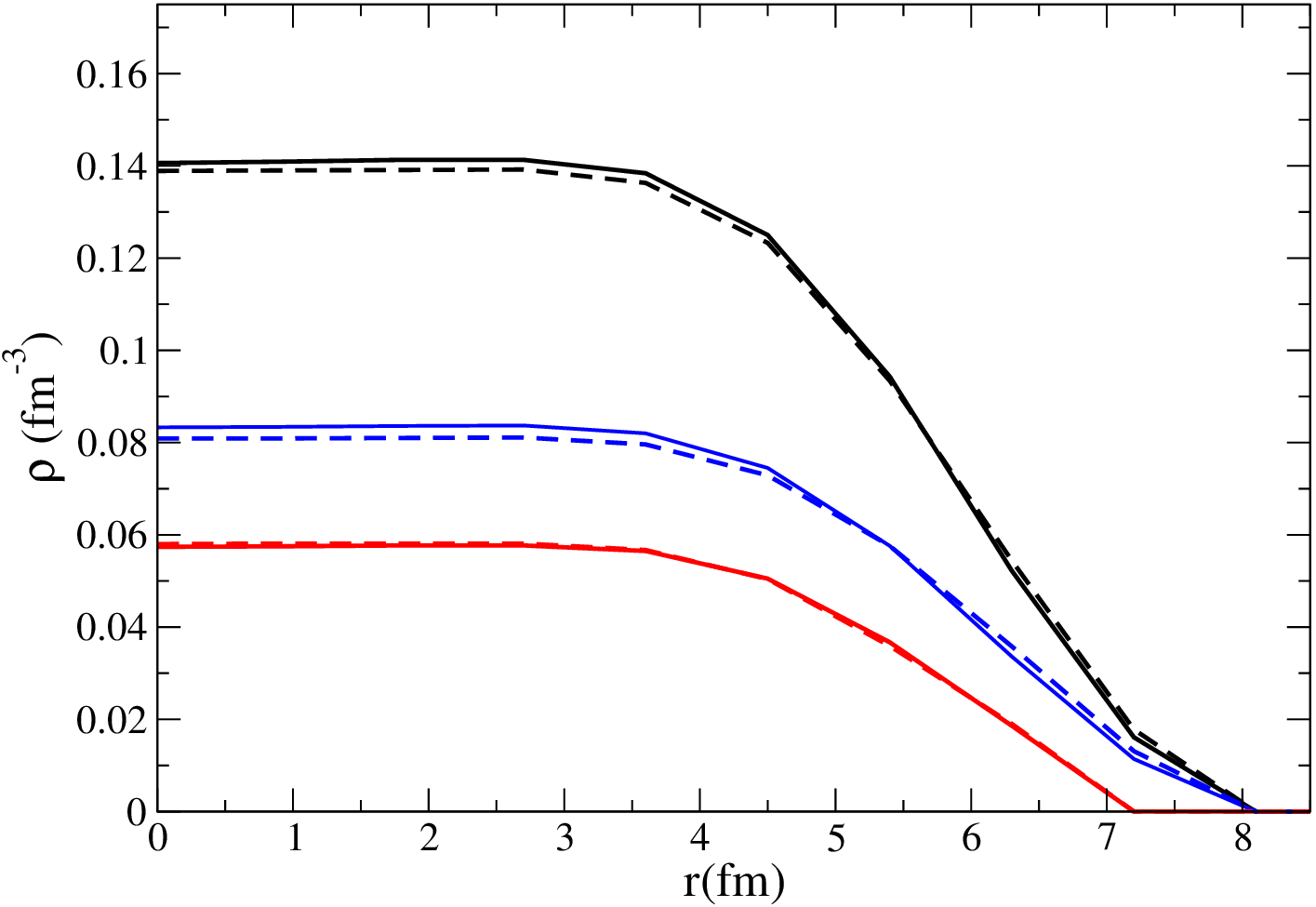}
\end{center}
\caption{(Color online) The total (black), neutrons (blue) and
protons (red) radial density profiles for asysoft (solid lines) and asysuperstiff (dashed lines).}
\label{densprof}
\end{figure}

The numerical procedure to integrate the transport equations is based on the
test-particle (t.p.) method. For a good spanning of phase-space we work with $1200$ t.p. per nucleon.
We consider the
neutron rich nucleus $^{132}Sn$ and we
determine its ground state configuration as the equilibrium (static)
solution of Eq.(\ref{vlasov}). Then  proton and neutron densities
$\displaystyle \rho_q(\vec{r},t)=\int \frac{2 d^3 {\bf p}}{(2\pi\hbar)^3}f_q(\vec{r},\vec{p},t)$
can be evaluated.
The radial density profiles
for two asy-EOS
are reported in Fig. \ref{densprof}.
As an additional check of our initialization procedure,
the neutron and proton mean square radii
$\displaystyle \langle r_q^2 \rangle = \frac{1}{N_q} \int r^2 \rho_q(\vec{r},t) d^3 {\bf r}$,
as well as the skin thickness
$\displaystyle \Delta R_{np}= \sqrt{\langle r_n^2 \rangle}-\sqrt{\langle r_p^2 \rangle}$,
were also calculated in the ground state and shown in Table \ref{table1}.
\begin{table}
\begin{center}
\begin{tabular}{|l|r|r|r|r|r|} \hline
asy-EoS       & $E_{sym}/A$    & L(MeV)  & $R_n$(fm) & $R_p$(fm) & $\Delta R_{np}(fm)$  \\ \hline
asysoft       &     30.                  & 14.     & 4.90  & 4.65    & 0.25 \\ \hline
asystiff      &     28.                  & 73.     & 4.95  &  4.65   & 0.30 \\ \hline
asysupstiff   &     28.                  & 97.     & 4.96  &  4.65   & 0.31 \\ \hline
\end{tabular}
\caption{The symmetry energy at saturation (in MeV), the slope parameters, neutron rms radius,
protons rms radius, neutron skin thickness for the three asy-EoS.}
\label{table1}
\end{center}
\end{table}
The values obtained with our semi-classical approach
are in a reasonable agreement with those reported by employing other
models for similar interactions \cite{paar2005}.
The neutron skin thickness is increasing with the slope parameter,
as expected from a faster reduction of the symmetry term on the surface \cite{yoshida2004,baran2005}.
This feature has been discussed in detail in \cite{carbone2010}.

\begin{figure}
\begin{center}
\includegraphics*[scale=0.32]{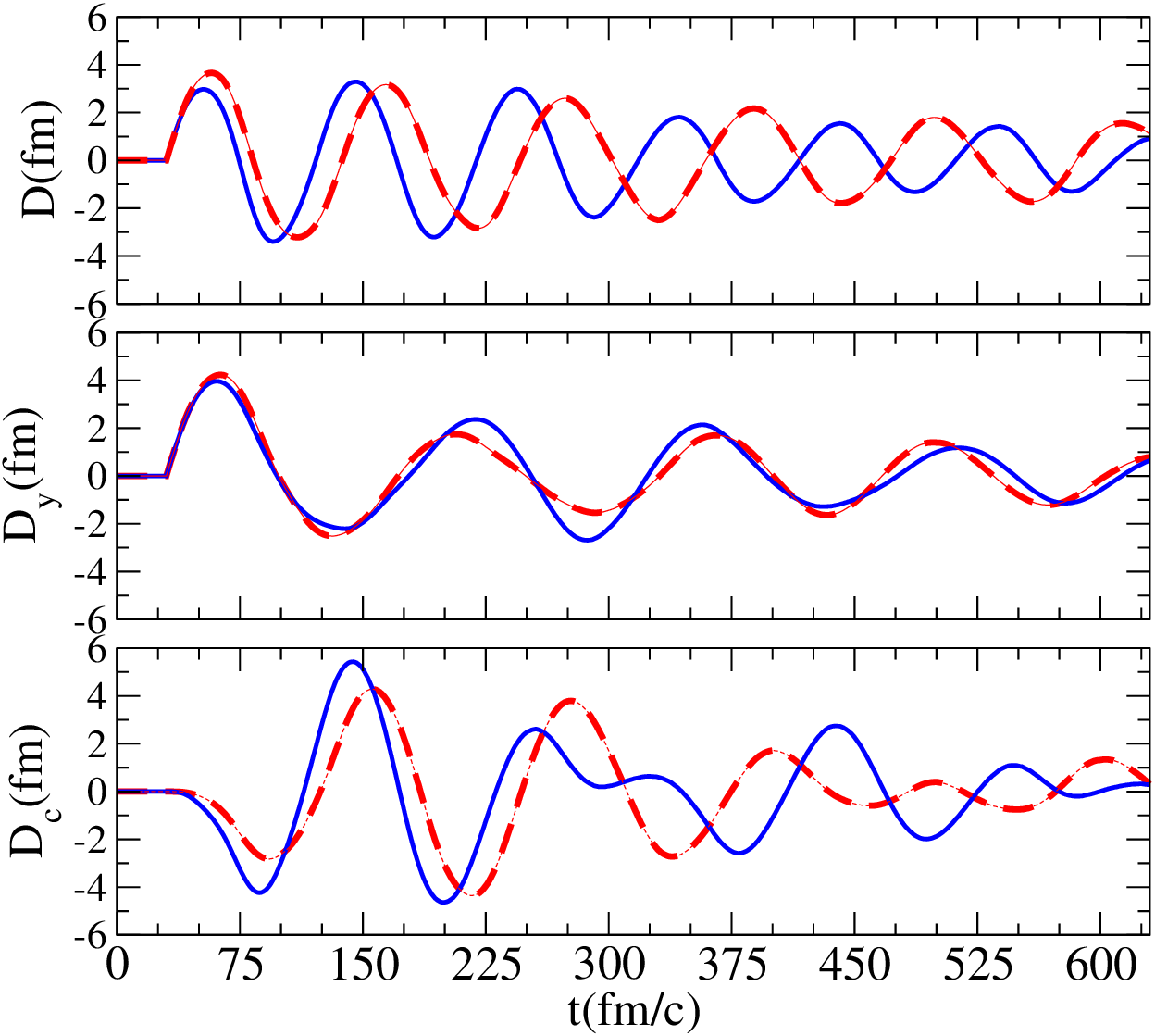}
\end{center}
\caption{(Color online) The time evolution of the total dipole $D$ (top)
of the dipole $D_y$ (middle) and of core dipole $D_c$
for asysoft (blue, solid) and asysuperstiff (red, dashed) EoS. Pygmy-like initial excitation.}
\label{diptime}
\end{figure}

  To inquire on the collective properties of the pygmy dipole we excite the nuclear system
at the initial time $t=t_0 = 30fm/c$ by boosting along the $z$
direction all excess neutrons and in opposite direction all core
nucleons, while keeping the CM of the nucleus at rest (Pygmy-like
initial conditions).
The excess neutrons were identified as the most
distant $N_e=32$ neutrons from the nucleus CM. Then the system is left to
evolve and the evolution of the collective coordinates $Y$, $X_c$ and $X$,
associated with the  different dipole modes,
is followed for $600 fm/c$
by solving numerically the equations (\ref{vlasov}).
The simple estimate of the EWSR provided
by the HOSM suggests, when compared to experiments, 
that some of the $N_e$ neutrons boosted
in the initial conditions are still bound to the core.
This is confirmed by the transport simulations.  
Indeed, apart from the quite undamped oscillations of the $Y$ coordinate, 
we also remark that the core does not remain inert.
In Fig. \ref{diptime} we plot the time evolution of the dipole
$D_y$, of the total dipole $D$ and core dipole $D_c$ moments, for two asy-EoS.
As observed, 
while $D_y$ approaches its maximum
value, an oscillatory motion of the dipole $D_c$ initiates and
this response is symmetry energy dependent: larger is the slope
parameter $L$, more delayed is the isovector core reaction.
 This can be explained in terms of low-density (surface)
contributions to the vibration and therefore of the density behavior
of the symmetry energy below normal density: a larger L corresponds
to a larger neutron presence in the surface and so to a smaller
coupling to the core protons.
We see that the total dipole $D(t)$ is strongly affected by the presence of
isovector core oscillations,
mostly related to the isovector part of the effective interaction.
Indeed, $D(t)$ gets a higher oscillation frequency with respect to
$D_y$, sensitive to the asy-EOS. The fastest
vibrations are observed in the asysoft case, which gives the largest
value of the symmetry energy below saturation. In correspondence the
frequency of the pygmy mode seems to be not much affected by the
trend of the symmetry energy below saturation, see also next Fig.
\ref{dipspectrum}, clearly showing the different nature,
isoscalar-like, of this oscillation. For each case we calculate the
power spectrum of $D_y$:
$\displaystyle |D_y (\omega)| ^2 = |\int_{t_0}^{t_{max}} D_y(t) e^{-i\omega t} dt|^2$
and similarly for $D$. The results are shown in Fig.
\ref{dipspectrum}. The position of the centroids corresponding to
the GDR shifts toward larger values when we move from superasystiff (largest slope parameter $L$)
to asysoft EoS.
This evidences the importance of the volume, S-J component of the GDR in
 $^{132}Sn$.
The energy centroid associated with the PDR is situated below the GDR peak, at around $8.5 MeV$,
quite insensitive to the asy-EOS,
pointing to an isoscalar-like nature of this mode.
A similar conclusion
was reported within a relativistic mean-field approach
\cite{liang2007}. While in the schematic HOSM all
dipole modes are degenerate, with an energy $E=41 A^{-1/3} \approx
8MeV$ for $^{132}Sn$, within the Vlasov approach the GDR energy is
pushed up by the isovector interaction. Hence the structure of the
dipole response can be explained in terms of the development of
isoscalar-like (PDR) and isovector-like (GDR) modes, as 
observed in asymmetric systems \cite{baran2001a}.
\begin{figure}
\begin{center}
\includegraphics*[scale=0.34]{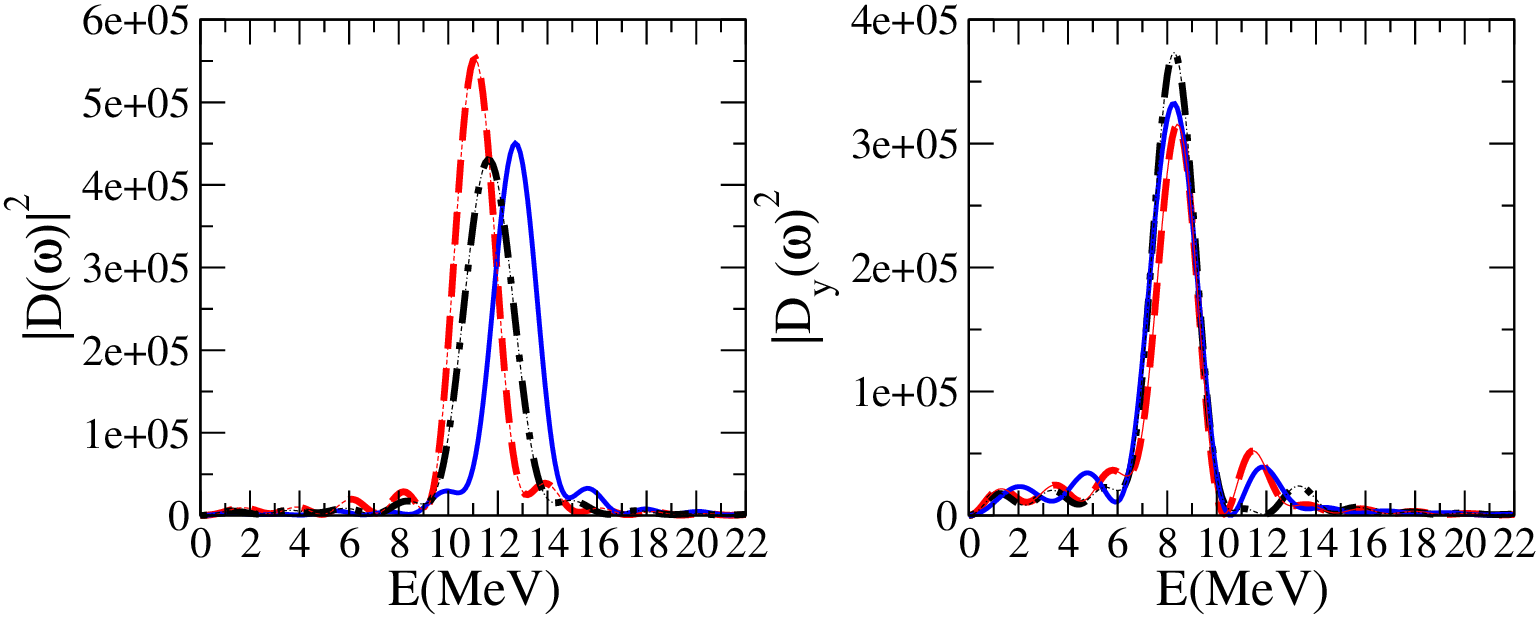}
\end{center}
\caption{(Color online) The power spectrum of total dipole (left) and
of  the dipole $D_y$ (right) (in $fm^4/c^2$), for asysoft
(blue, solid line), asystiff (black, dot-dashed line) and asysuperstiff (red, dashed line)
EoS. Pygmy-like initial conditions.
}
\label{dipspectrum}
\end{figure}
Both modes are excited in the considered, pygmy-like initial conditions. Looking
at the total dipole mode direction, that is close to the isovector-like normal mode,
one observes a quite large contribution in the GDR region. On the other hand, 
considering the $Y$ direction, more closely related to the isoscalar-like mode, a larger
response amplitude is detected in the pygmy region.
\begin{figure}
\begin{center}
\includegraphics*[scale=0.34]{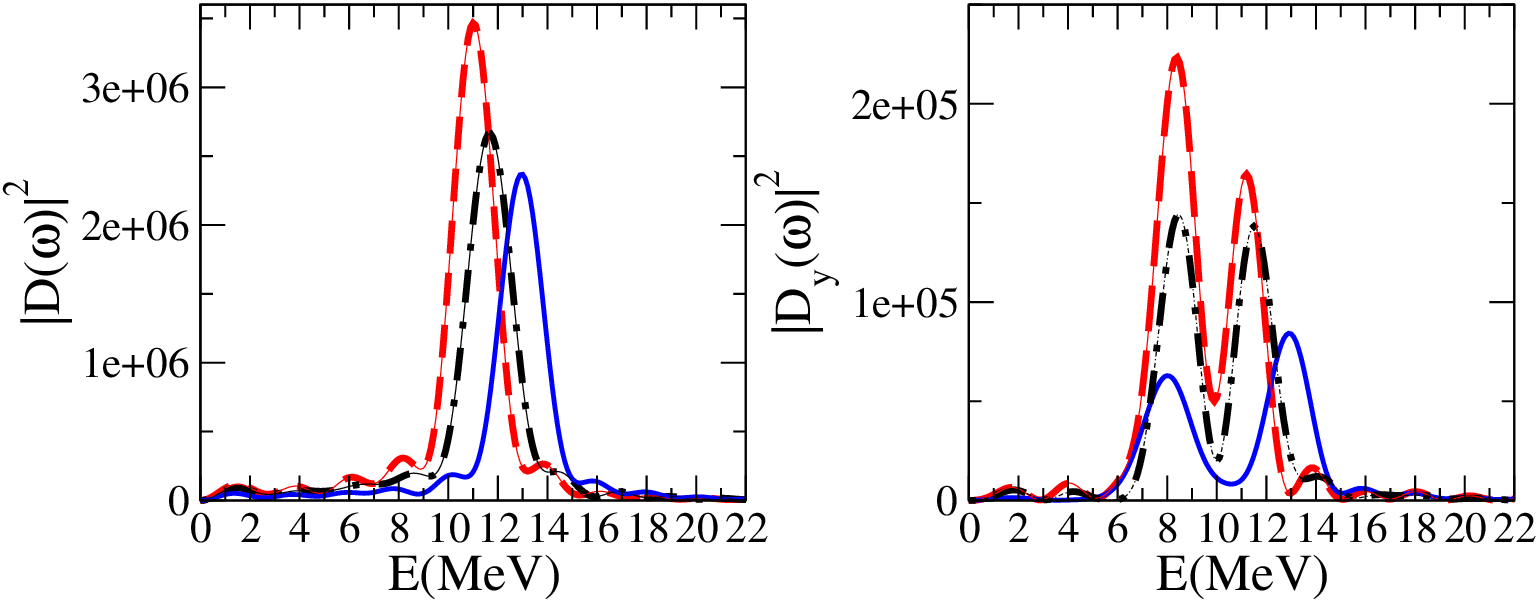}
\end{center}
\caption{(Color online) The same as in Fig. \ref{dipspectrum} but for a GDR-like
initial excitation.}
\label{gdrspectrum}
\end{figure}
To check the influence of the initial conditions on the dipole
response, let us consider the case of a  GDR-like excitation, corresponding
to  a boost of all neutrons against all protons, keeping the CM at rest.
The initial collective energy corresponds to first GDR excited state, around $15MeV$. 
Now the initial excitation favours the isovector-like mode and
even in the $Y$ direction we observe a sizeable contribution in the
GDR region, see the Fourier spectrum of $D_y$ in Fig. \ref{gdrspectrum}. 
From this result it clearly emerges that 
a part of the $N_e$ excess neutrons is involved in a GDR type motion
and the relative weight depends on the symmetry
energy: more neutrons 
are involved in the pygmy mode in the 
asysuperstiff EOS case, in connection to the larger neutron skin size.
We have also checked that, if the coordinate Y is constructed
by taking the $N_y$ most distant neutrons (with $N_y < N_e$),
the relative weight increases in the PDR region. 
In any case, since part of the excess nucleons 
contributes to the GDR mode, a lower EWSR value than 
the HOSM predictions corresponding to $N_y=N_e$ is expected. 
Indeed, in the Fourier power spectrum of $D$ in Fig.
\ref{gdrspectrum}, a weak response is seen at the pygmy frequency.

These investigations also rise the question of the appropriate way to
excite the PDR. Nuclear rather than electromagnetic probes
can induce neutron skin excitations closer to our first class of initial conditions
\cite{vitturi2010}.
In the case of the GDR-like initial excitation we can relate the strength function
to $Im(D(\omega))$ \cite{suraud1997} and then the corresponding
cross section can be calculated. Our estimate of the integrated cross section 
over the PDR region represents  $2.7 \%$
for asysoft, $4.4 \%$ for asystiff and $4.5 \%$ for asysuperstiff,
 out of the total cross section. Hence the EWSR
exhausted by the PDR is proportional to the skin thickness, in agreement
with the results of \cite{inakura2011}.
The fraction of photon emission probability associated with the PDR region
can be estimated from the total dipole acceleration, within a
bremsstrahlung approach \cite{baran2001}.
We obtain a percentage
of $4.7 \%$ for asysoft, $7.7 \%$ for asystiff and $9 \%$ for
asysuperstiff EOS, consistent with the previous interpretation.

 Summarizing, in this work we evidence, both within HOSM and a
semi-classical Landau-Vlasov approach, the existence, in neutron
rich nuclei, of a collective pygmy dipole mode determined by the
oscillations of some excess neutrons against the nuclear
core. From the transport simulation results the PDR energy centroid
for $^{132}Sn$ appears around $8.5$ $MeV$, rather insensitive to the
density dependence of the symmetry energy and well below the GDR peak.
This supports the isoscalar-like character of this collective
motion. A complex pattern, involving the coupling  of the neutron skin with
the core dipole mode, is noticed.
While HOSM can provide some predictions of the EWSR fraction exhausted
by the pygmy mode, $\displaystyle f_y=\frac{N_y Z}{N A_c}$, depending on the
number $N_y\leq N_e$ of neutrons involved, the transport model
indicates that part of the excess neutrons $N_e$ are coupled to the 
GDR mode and gives some hints about 
the number of neutrons, $N_y$, actually partecipating in the pygmy mode.
This reduces considerably the EWSR acquired by the PDR, our numerical
estimate providing values well below $10 \%$, but proportional to the
symmetry energy slope parameter $L$, that affects the number of
excess neutrons on the nuclear surface. We consider these effects as
related also to the S-J component of the dipole dynamics
in medium-heavy nuclei. It is therefore interesting to extend the
present analysis to lighter nuclei, like Ni or Ca isotopes, where
the Goldhaber-Teller component can be more important. We would like
to mention that such self-consistent, transport  approaches,
can be valuable in exploring the collective response of other
mesoscopic systems where similar normal modes may manifest, see
\cite{mesoscopic}.

 This work for V. Baran was supported by a grant of the Romanian National
Authority for Scientific Research, CNCS - UEFISCDI, project number PN-II-ID-PCE-2011-3-0972.
For B. Frecus this work was supported by the strategic grant POSDRU/88/1.5/S/56668.


\begin{thebibliography}{99}
\bibitem{ncore}
If naively we assume the core neutrons to be the ones of the most
bound Sn isotopes ($^{118}Sn$ and $^{120}Sn$) we get as excess
neutrons $N_e \simeq 12 \div 14$. The HOSM-EWSR fraction is then
reduced to $f_y \simeq 7.3 \div 8.5 \%$.
\end{thebibliography}
\end{document}